\documentstyle[12pt,fps]{article}
\newcommand{\Z}{{\sf Z \!\!\! Z}}

\newcommand{\vspu}{\vspace*{6mm}}
\newcommand{\mspa}{\, \,}

\newcommand\be{\begin{equation}}
\newcommand\ee{\end{equation}}
\newcommand\bea{\begin{eqnarray}}
\newcommand\eea{\end{eqnarray}}

\makeatletter
\@addtoreset{equation}{section}
\makeatother
\begin{document}
\begin{center}

{\LARGE Perfect Lattice Topology: \\ \vspu

The Quantum Rotor as a Test Case} 
\footnote{This work is supported in part by funds provided by the U.S.
Department of Energy (D.O.E.) under cooperative research agreement
DE-FC02-94ER40818.}
\vspu
\vspu

W. Bietenholz$^{\rm a}$, R. Brower$^{\rm b}$, 
S. Chandrasekharan$^{\rm c}$ and U.-J. Wiese$^{\rm c}$\\
\vspu

$^{\rm a}$ HLRZ c/o Forschungszentrum, J\"{u}lich \\
52425 J\"{u}lich, Germany \\
\vspu

$^{\rm b}$ Department of Physics \\
Boston University \\
Boston MA 02215, USA \\
\vspu

$^{\rm c}$ Center for Theoretical Physics \\
Laboratory for Nuclear Science and Department of Physics \\
Massachusetts Institute of Technology \\
Cambridge MA 02139, USA \\
\vspu
Preprint \ MIT-CTP 2583, HLRZ 01/97
\end{center}
\vspu
\vspu 

 
Lattice actions and topological charges that are classically and
quantum mechanically perfect (i.e. free of lattice artifacts) are
constructed analytically for the quantum rotor. It is demonstrated
that the Manton action is classically perfect while the Villain action
is quantum perfect.  The geometric construction for the topological
charge is only perfect at the classical level. The quantum perfect
lattice topology associates a topological charge distribution, not
just a single charge, with each lattice field configuration.  For the
quantum rotor with the classically perfect action and topological
charge, the remaining cut-off effects are exponentially suppressed.
 
 
\newpage

\noindent{\bf Introduction}

The most severe systematic errors in numerical simulations of lattice field
theories are due to finite lattice spacing effects. For asymptotically free 
theories like QCD, Hasenfratz and Niedermayer \cite{Has94} realized that in the 
classical limit one can numerically construct nonperturbative perfect lattice 
actions, which are completely free of finite lattice spacing 
artifacts. Using the
classically perfect action of the 2-d $O(3)$ model even in the quantum theory, 
they found that cut-off effects are still practically eliminated. A similar
behavior has been observed in the 4-d pure $SU(3)$ gauge theory \cite{DeG95},
and it has been argued that in these cases the classically perfect action
is also quantum perfect at the 1-loop level. This has been demonstrated 
explicitly
for the 2-d $O(3)$ model \cite{Far95}. In the Gross-Neveu model at large $N$, 
the classically perfect action is also perfect at the quantum level for
arbitrarily short correlation lengths \cite{Bie95}. 
The same is true for the $O(N)$ model at large $N$. Still,
one expects that in general a classically perfect action will not be quantum
perfect. In this paper we investigate this question in a simple model ---
the quantum rotor (or 1-d $XY$ model). In this case both, the classically and
quantum perfect actions can be constructed analytically. In particular, when 
the classically perfect action is used at the quantum level, the size of the 
remaining cut-off effects can be analyzed in detail.

The topological charge of a field configuration on the lattice is
usually defined by some smooth interpolation. If one uses the standard
lattice formulation, the artifacts are notoriously large.
A better definition was proposed by Berg and L\"{u}scher
\cite{Lusch} and is referred to as the geometric method.  This
definition has the virtue that the topological charge is always an
integer.
However, artifacts can still be significant
for rough configurations. 
Here we use the quantum rotor to formulate
a perfect prescription for the topology of lattice
configurations. This can be done explicitly since the model is exactly
solvable. 

The quantum rotor is an ideal test case for perfect lattice
topology. One can introduce a topological charge and a $\theta$-vacuum
parameter in analogy to QCD. In the 2-d $O(3)$ model the classically
perfect topological charge has been constructed in Ref.\cite{Bla96}.
Similar constructions have been carried out in the 2-d $CP^3$ model
\cite{Bur95} and in 4-d pure $SU(2)$ gauge theory \cite{DeG96}. Again,
the classically perfect topological charge is not
perfect at the quantum level. For the quantum rotor the lattice
topological charge can be constructed explicitly both at the classical
and at the quantum level. This allows us to study the size of the
remaining cut-off effects when the classically perfect topological
charge is used in the quantum theory. We don't expect the cut-off
effects of the quantum rotor to be in quantitative agreement with
those of quantum field theory models in higher dimensions. Still, at
a qualitative level the analytic study of the quantum rotor tells us
how lattice artifacts depend on the correlation length.

The study of the quantum rotor also sheds light on the structure of
quantum perfect lattice topology. It turns out that at the quantum
level a lattice field configuration is not just characterized by a
single topological charge but by a whole topological charge
distribution. This insight may lead to new ways of approximating
perfect lattice topology in physically relevant models like QCD.

\noindent{\bf Transfer Matrix for the Quantum Rotor}

We consider a particle of mass $M$ confined to a circle of radius
$R$. This system represents a rotor with moment of inertia $I = M
R^2$. The corresponding Hamilton operator reads
\begin{equation}
H = - \frac{\hbar^2}{2 I} \partial_\varphi^2 \ ,
\end{equation}
where $\varphi$ is an angle describing 
the position of the particle. The
energy spectrum of the quantum rotor is given by
\begin{equation}
E_m = \frac{\hbar^2 m^2}{2 I},
\end{equation}
where $m \in \Z$ specifies the angular momentum. The corresponding wave
functions,
\begin{equation}
\label{eq:eigenvec}
\langle \varphi|m\rangle = \frac{1}{\sqrt{2 \pi}} \exp \{ i m \varphi \} \; ,
\end{equation}
are $2\pi$ periodic in $\varphi$.
The above quantum rotor can be modified by changing the Hamilton operator to
\begin{equation}
H(\theta) = - \frac{\hbar^2}{2 I} \Big( \partial_\varphi 
 - i \frac{\theta}{2 \pi}\Big) ^2.
\end{equation}
The modified rotor has nontrivial topological properties. 
The angle $\theta$ is 
analogous to the vacuum angle in QCD. The energy eigenvalues of the modified 
quantum rotor are
\begin{equation}
E_m(\theta) = \frac{\hbar^2}{2 I} \Big( 
m - \frac{\theta}{2 \pi} \Big) ^2,
\end{equation}
while the wave functions remain unchanged. The partition function for the 
modified quantum rotor at finite temperature $T = 1/\beta$ is given by
\begin{equation}
Z(\theta) = \mbox{Tr} \mspa \exp \{ - \beta H(\theta)\} 
= \sum_{m \in \Z} 
\exp \{ - \frac{\beta \hbar^2}{2 I} (m - \frac{\theta}{2 \pi})^2 \} .
\end{equation}
From the partition function one derives the topological charge distribution,
\begin{equation}
p(Q)\;=\;\frac{1}{2 \pi} \int_{-\pi}^\pi d\theta \ Z(
\theta) 
\exp \{ - i \theta Q \}\;=\;
\sqrt{\frac{2 \pi I}{\beta \hbar^2}} 
\exp \{ - \frac{2 \pi^2 I}{\beta \hbar^2} Q^2 \} ,
\end{equation}
and from that the topological susceptibility
\begin{equation}
\chi_t = \frac{1}{\beta} \frac{\sum_{Q \in \Z}\; Q^2 p(Q)}
{\sum_{Q \in \Z} \;p(Q)}
= \frac{1}{\beta} 
\frac{\sum_{Q \in \Z}\; Q^2 \exp \{ - 2 \pi^2 I Q^2/\beta \hbar^2 \} }
{\sum_{Q \in \Z}\; \exp \{ - 2 \pi^2 I Q^2/\beta \hbar^2 \} }.
\end{equation}
In the zero temperature limit $\beta \rightarrow \infty$ one obtains
\begin{equation}
\label{topsus}
\chi_t = \frac{d^2 E_0(\theta)}{d \theta^2}|_{\theta = 0} = 
\frac{\hbar^2}{4 \pi^2 I}.
\end{equation}
Using the energy gap at $\theta = 0$,
we introduce a correlation time (correlation length in Euclidean time)
\begin{equation}
\xi = \frac{\hbar}{E_1 - E_0} = \frac{2 I}{\hbar},
\end{equation}
such that at zero temperature
\begin{equation}
\frac{\chi_t \; \xi}{\hbar} = \frac{1}{2 \pi^2}.
\end{equation}

To formulate the path integral for the quantum rotor on a Euclidean time 
lattice with lattice spacing $a$, we first construct the transfer matrix,
\begin{eqnarray}
\label{tmatrix}
\langle \varphi_{t+a}|{\cal T}|\varphi_t \rangle&=&
\langle \varphi_{t+a}|\exp \{ - \frac{a}{\hbar} H(\theta) \}
|\varphi_t \rangle \nonumber \\
&=&\sum_{m \in \Z} \langle \varphi_{t+a}|m \rangle 
\exp \{ - \frac{\hbar a}{2 I} (m - \frac{\theta}{2 \pi})^2 \}
\langle m|\varphi_t \rangle \nonumber \\
&=&\frac{1}{2 \pi} \sum_{m \in \Z} 
\exp \{ - \frac{\hbar a}{2 I} (m - \frac{\theta}{2 \pi})^2
+ i m (\varphi_{t+a} - \varphi_t) \} .
\end{eqnarray}
Hence, the Fourier transform of the transfer matrix with respect to
$\varphi = \varphi_{t+a} - \varphi_t$ is 
$\exp \{ - \hbar a (m - \theta/2 \pi)^2/2 I \} /2 \pi$. 
Using the 
Poisson re-summation formula, the transfer matrix can also be written as,
\begin{equation} \label{transfer}
\langle \varphi_{t+a}|{\cal T}|\varphi_t \rangle =
\sqrt{\frac{I}{2 \pi \hbar a}} \sum_{n \in \Z} \exp \{ 
- \frac{I}{2 \hbar a}
(\varphi_{t+a} - \varphi_t + 2 \pi n)^2 
+ i \frac{\theta}{2 \pi} (\varphi_{t+a} - \varphi_t + 2 \pi n)\} .
\end{equation}
This follows when we consider the Fourier transform of this expression
with respect to $\varphi$,
\begin{eqnarray}
&&\frac{1}{2 \pi}
\int_{-\pi}^\pi d\varphi \ \sqrt{\frac{I}{2 \pi \hbar a}} \sum_{n \in \Z} 
\exp \{ - \frac{I}{2 \hbar a} (\varphi + 2 \pi n)^2 + i \frac{\theta}{2 \pi} 
(\varphi + 2 \pi n) - i m \varphi \} \nonumber \\
&&=\frac{1}{2 \pi} \exp \{ 
- \frac{\hbar a}{2 I} (m - \frac{\theta}{2 \pi})^2 \} ,
\end{eqnarray}
which agrees with the Fourier transform of the expression
in Eq.(\ref{tmatrix}).

The exact partition function can be written as a path integral on a
Euclidean time lattice with $N = \hbar \beta/a$ points. Starting from
\begin{equation}
Z = \mbox{Tr} \mspa \exp \{ - \beta H \} = \mbox{Tr} \ {\cal T}^N,
\end{equation}
one inserts complete sets of states $|\varphi_t\rangle$ at each time
step (Chapman-Kolmogoroff equation) and one obtains
\begin{equation} \label{Villpart1}
\label{eq:QuantumPerf}
Z = \int {\cal D}\varphi {\cal D}n \ \exp \{ - \frac{1}{\hbar} S[\varphi,n]
+ i \theta Q[\varphi,n] \} .
\end{equation}
The measure of the path integral is given by
\begin{equation} \label{Villpart2}
\int {\cal D}\varphi {\cal D}n = \prod_{t=0}^{(N-1)a} 
\sqrt{\frac{I}{2 \pi \hbar a}} \int_{-\pi}^\pi d\varphi_t 
\sum_{n_{t+a/2} \in \Z} \ ,
\end{equation}
with periodic boundary conditions $\varphi_{Na} = \varphi_0$. The action takes 
the form
\begin{equation} \label{Villpart3}
S[\varphi,n] = a \sum_{t=0}^{(N-1)a} \frac{I}{2}
\Big( \frac{\varphi_{t+a} - \varphi_t 
+ 2 \pi n_{t+a/2}}{a} \Big) ^2.
\end{equation}
This is the Villain action of the $XY$ model, which --- by construction
--- turns out to be a {\em quantum perfect action}. In the continuum
limit $a \rightarrow 0$, it converges to the continuum action
\begin{equation}
S[\varphi] = \int_0^{\hbar \beta} dt \ \frac{I}{2} \dot \varphi(t)^{2}  \; .
\end{equation}
Similarly, the perfect topological charge is given by
\begin{equation} \label{PerfTopch}
Q[\varphi,n] = \frac{a}{2 \pi} \sum_{t=0}^{(N-1)a} 
\frac{\varphi_{t+a} - \varphi_t + 2 \pi n_{t+a/2}}{a},
\end{equation}
which --- due to periodicity in time --- is equivalent to
$Q[\varphi,n] =  \sum_{t=0}^{(N-1)a} n_{t+a/2}$.
In the continuum limit it turns into
\begin{equation} \label{Qphidot}
Q[\varphi] = \frac{1}{2 \pi}\int_0^{\hbar \beta} dt \ \dot \varphi (t).
\end{equation}

It is interesting that the quantum perfect topological charge is given
in terms of both, $\varphi$ and $n$.  This is in contrast to familiar
lattice definitions of the topological charge, which work with the
angles alone.  If one insists to represent also the perfect
topological charge in terms of $\varphi$ alone, one has to perform the
sums over $n$. This, however, does not result in a single topological
charge associated with a configuration $[\varphi]$. Instead, even a
single configuration is associated with a whole topological charge
distribution
\begin{equation}
p[\varphi](Q) = \prod_{t=0}^{(N-1)a} \sum_{n_{t+a/2} \in \Z}
\exp \{ - \frac{1}{\hbar} S[\varphi,n]\} \delta_{Q,Q[\varphi,n]},
\end{equation}
and the total topological charge distribution is given by
\begin{equation}
p(Q) = \int {\cal D}\varphi \ p[\varphi](Q).
\end{equation}
This suggests a new way of thinking about lattice topology. Instead of 
associating a quantum field configuration with a single topological charge,
one should associate it with a whole charge distribution. In fact, one may 
think of a lattice field configuration as representing an ensemble of 
continuum field configurations that turn into the lattice configuration under
a renormalization group transformation. Then the topology of the continuum
theory is perfectly represented by the lattice field, if one associates with it
the topological charge distribution of the corresponding ensemble of continuum
configurations. 
This is exactly what happens for the quantum perfect topological charge.
As we will see below, one can explicitly construct renormalization 
group transformations for which the above perfect action lies on the 
renormalized trajectory.

Before that let us consider the classical limit $\hbar \rightarrow 0$. Then the
path integral turns into a saddle point problem. In particular, for a given
configuration $\varphi$ the variables $n$ adjust themselves such that
$(\varphi_{t+a} - \varphi_t + 2 \pi n_{t+a/2})^2$ is minimized, i.e. such 
that $\varphi_{t+a} - \varphi_t + 2 \pi n_{t+a/2} = 
(\varphi_{t+a} - \varphi_t) \mspa \mbox{mod} \mspa
2 \pi \in ]-\pi,\pi]$. Hence, in
the classical limit the perfect action turns into
\begin{equation}
S_c[\varphi] = \sum_{t=0}^{(N-1)a} \frac{I}{2 a}
((\varphi_{t+a} - \varphi_t) \mspa \mbox{mod} \mspa 2 \pi)^2.
\end{equation}
This is the so-called Manton action \cite{Manton},
which turns out to be classically
perfect. Similarly, the classically perfect topological charge is
\begin{equation}
Q_c[\varphi] = \frac{1}{2 \pi} \sum_{t=0}^{(N-1)a}
(\varphi_{t+a} - \varphi_t) \mspa \mbox{mod} \mspa 2 \pi.
\end{equation}
This is identical with the geometric topological charge. It is classically
but not quantum perfect. The geometric charge is based on a particular 
interpolation of the lattice variables $\varphi_t$ along a shortest arc. The
quantum perfect charge, on the other hand, emerges when one integrates over
all possible interpolations (all paths connecting neighboring lattice
variables) with the appropriate Boltzmann weight.

\noindent{\bf Comparison between Classical and Quantum Perfection}

The question arises, how well the classically perfect action and topological
charge work at the quantum level. This is an important issue, because
in the more complicated models in higher dimensions, we only have the 
classically perfect quantities at hand. To investigate this question we 
construct the transfer matrix for the classically perfect action
\begin{eqnarray}
\langle \varphi_{t+a}|{\cal T}_c|\varphi_t \rangle&=&
\sqrt{\frac{I}{2 \pi \hbar a}} 
\exp \{ - \frac{I}{2 \hbar a} ((\varphi_{t+a} - \varphi_t) 
\mspa \mbox{mod} \mspa 2 \pi)^2 \}
\nonumber \\
&\times&\exp \{ i \frac{\theta}{2 \pi} (\varphi_{t+a} - \varphi_t) 
\mspa \mbox{mod} \mspa 2 \pi \} .
\end{eqnarray}
It is straightforward to diagonalize the transfer matrix by performing
a Fourier transform. The resulting energy spectrum is given by
\begin{equation}
\exp \{ - \frac{a}{\hbar} E_{m}^{c}(\theta) \}
= \sqrt{\frac{I}{2 \pi \hbar a}}
\int_{-\pi}^\pi d\varphi \ \exp \{ - \frac{I}{2 \hbar a} \varphi^2
- i(m - \frac{\theta}{2 \pi})\varphi \} .
\end{equation}
For the topological susceptibility at zero temperature this implies
\begin{equation}
\chi^c_t = \frac{d^2 E_{0}^{c}(\theta)}{d\theta^2}|_{\theta = 0} =
\frac{\hbar}{4 \pi^2 a} \frac{\int_{-\pi}^\pi d\varphi \ \varphi^2
\exp \{ - I \varphi^2/2 \hbar a \} }{\int_{-\pi}^\pi d\varphi \ 
\exp \{ - I \varphi^2/2 \hbar a \} }.
\end{equation}
Of course, in the continuum limit $a \rightarrow 0$ one reproduces the
result of Eq.(\ref{topsus}). At large $\xi /a$, the lattice artifacts
are exponentially suppressed as
\begin{equation}
\frac{\chi^{c}_{t} \; \xi}{\hbar} \sim 
\frac{1}{2\pi^{2}} - \sqrt{\frac{\pi \xi}{a}}
\; \exp \{ - \frac{\pi^2 \xi}{4a} \} .
\end{equation}

For comparison we also consider the standard action
\begin{equation}
S_s[\varphi] = \sum_{t=0}^{(N-1)a} \frac{I}{a}
(1 - \cos(\varphi_{t+a} - \varphi_t)).
\end{equation}
Using it together with the geometric topological charge one finds
\begin{eqnarray}
\langle \varphi_{t+a}|{\cal T}_s|\varphi_t \rangle&=&
\sqrt{\frac{I}{2 \pi \hbar a}}
\exp \{ - \frac{I}{\hbar a} (1 - \cos(\varphi_{t+a} - \varphi_t)) \}
\nonumber \\
&\times&\exp \{ i \frac{\theta}{2 \pi} (\varphi_{t+a} - \varphi_t)
\mspa \mbox{mod} \mspa 2 \pi \} .
\end{eqnarray}
Again we diagonalize the transfer matrix by performing a Fourier transformation
and we arrive at
\begin{equation}
\exp\{- \frac{a}{\hbar} E_{m}^{s}(\theta)\} = \sqrt{\frac{I}{2 \pi \hbar a}}
\int_{-\pi}^\pi d\varphi \ \exp \{ - \frac{I}{\hbar a} (1 - \cos\varphi)
- i(m - \frac{\theta}{2 \pi})\varphi \} .
\end{equation}
Now the topological susceptibility at zero temperature takes the form
\begin{eqnarray}
\chi_{t}^{s} &=& 
\frac{\hbar}{4 \pi^2 a} \frac{\int_{-\pi}^\pi d\varphi \ \varphi^2
\exp \{ - I (1 - \cos\varphi)/\hbar a \} }{\int_{-\pi}^\pi d\varphi \
\exp \{ - I (1 - \cos\varphi)/\hbar a \} } , \\
\frac{\chi_{t}^{s} \; \xi}{\hbar} &=& \frac{1}{2\pi^{2}} \Big( 1 +
\frac{a}{\xi} \Big) + O ((a/\xi )^{2}) . \label{linart}
\end{eqnarray}

Let us compare the magnitude of the 
cut-off effects for the various actions.
Fig.1 shows the ratio of the first two energy gaps 
$(E_2 - E_0)/(E_1 - E_0)$ at $\theta = 0$ as a function of
the correlation time in lattice units, $\xi/a$. 
\begin{figure}[hbt]
\hspace*{1cm}
\def\fpsangle{0}
\epsfxsize=120mm
\fpsbox{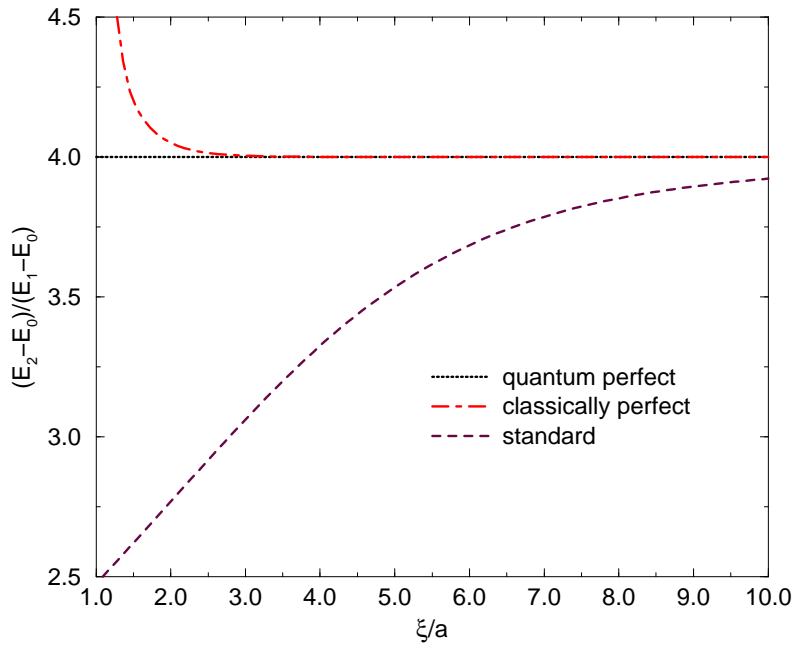}
\caption{\it The ratio of the first two energy gaps
$(E_2 - E_0)/(E_1 - E_0)$ as a function of the correlation time
in lattice units for the classically and quantum perfect actions and
for the standard action.}
\end{figure}
In the continuum limit
all actions give the correct answer. At finite $a$, however, both the 
standard and the classically perfect action suffer from 
cut-off effects. The asymptotic suppression of the artifacts
at large $\xi /a$ is again power-like respectively exponential,
\begin{eqnarray}
\frac{E_{2}^{s}-E_{0}^{s}}{E_{1}^{s}-E_{0}^{s}} &=&
4 \Big( 1 - \frac{a^{2}}{\xi^{2}} \Big) + O((a /\xi )^{3}), \\
\frac{E_{2}^{c}-E_{0}^{c}}{E_{1}^{c}-E_{0}^{c}} &=&
4 \Big( 1 + \frac{2}{\pi} \sqrt{\frac{\xi}{a}} \; \exp \{ -
\frac{\pi^{2} \xi}{4 a} + \frac{a}{\xi} \} + \dots \Big).
\end{eqnarray}
At a given ratio $\xi /a$, the artifacts are much smaller for
the classically perfect formulation than for the standard formulation.
For example, at $\xi = 2a$ the cut-off effect of the energy 
ratio is 30.8 percent for the standard action and only 1.3 percent 
for the classically perfect action.
\begin{figure}[hbt]
\hspace*{1cm}
\def\fpsangle{0}
\epsfxsize=120mm
\fpsbox{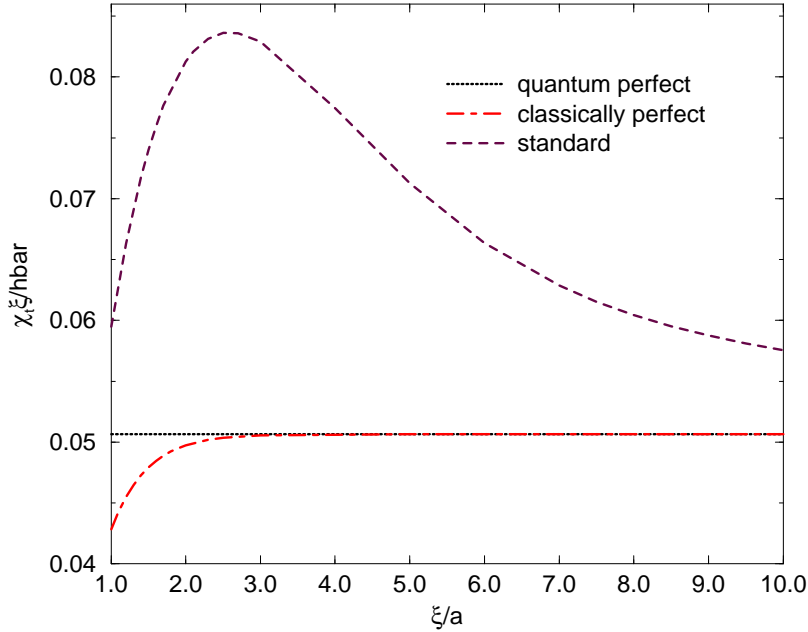}
\caption{\it The product $\chi_t \; \xi/\hbar$ 
as a function of the correlation 
time $\xi/a$ in lattice units for the classically and quantum perfect actions 
and for the standard action.}
\end{figure}
Fig.2 depicts the cut-off effect of $\chi_t \; \xi/\hbar$
at zero temperature as a function of $\xi/a$. 
Again, at $\xi = 2a$ the cut-off 
effect for the standard action and geometric charge is huge, 60.4 percent
to be precise, compared to 1.8 percent for the classically perfect action 
and topological charge. In this case the
artifacts in the standard action are even larger than in the
case of the gap ratio, because their suppression is only linear,
as we see from Eq.(\ref{linart}).

\noindent{\bf Quantum Perfect Action and the Renormalization Group}

Here we briefly present a more general real space renormalization
group derivation of the quantum perfect action for the quantum
rotor. This alternative derivation is interesting because it does not
rely explicitly on the peculiar simplifications inherent in a 1-d
theory and the resulting nearest neighbor form for the perfect
action. Above we have made use of the Hamilton operator (or
transfer matrix), which is especially simple for a 1-d system
since its dynamics is equivalent to ordinary quantum mechanics. 
In the 1-d case, by
integrating the fields on the line segments between each pair of lattice
points, one necessarily arrives at a nearest neighbor (or ultra-local)
action. Viewed as a renormalization group transformation (RGT),
such a procedure is often referred to as ``decimation''.  However, in
higher dimensions decimation is not a satisfactory blocking procedure
since it leads to long range interactions that are not exponentially
suppressed.  Consequently, it is important to ask if our results for
the quantum rotor and the qualitative lessons drawn from them are tied
to the use of the transfer matrix and ultra-locality.

A more general derivation can be made using a real space RGT for the
action in the path integral formalism.  In this form, the resulting
lattice action need not be ultra-local to formulate the quantum
perfect charge. Moreover such RGTs have recently been
used successfully to define classically perfect actions in higher
dimensions for certain quantum field theories~\cite{Schwing,QuaGlu} and
consequently they could in principle be used to investigate quantum
perfect topological features along the lines of this paper.  We are aware
of the difficulty in carrying out such a task in practice, but it is
still important that the quantum rotor example can also be understood
in this more general framework.

Let us sketch the derivation of this result. We divide the interval 
$[0,\hbar\beta]$ into $N$ parts, each with length $a$, so that 
$\hbar\beta = aN$. The blocked lattice field, $\phi$, is related to the 
continuum field, $\varphi$, by averaging over a cell of length $a$ 
centered at the lattice site $t$. Thus
\begin{equation}
\label{RGTcon}
\phi_t = \frac{1}{a} \int_{t - a/2}^{t + a/2} 
dt^\prime \; \varphi(t^\prime) \equiv  \int dt^\prime \; 
\Pi(t^\prime - t) \varphi (t^\prime) \; ,
\end{equation}
where the Haar function is $\Pi(t)= 1/a$ inside the cell $t \in
[-a/2,a/2]$ and zero outside.  In contrast, the transfer matrix
approach uses the value at the lattice site, which
corresponds to setting $\Pi(t^\prime - t) = \delta(t^\prime - t)$ in
Eq.(\ref{RGTcon}).  In fact, there is one {\em crucial} detail
that has been ignored in the definition of the blocking
prescription. Since the action is really a function of the compact
fields, $u(t) = \exp\{i \varphi(t)\}$, this blocking procedure must be
applied mod $2 \pi$. To define the perfect action, it will also be
convenient to generalize the blocking procedure  with Gaussian smearing.

Thus an exact perfect action can be defined by
\begin{equation}
\label{eq:FixPoint}
\exp \{ - \frac{1}{\hbar} S[\phi] \} = \int {\cal D}\varphi \exp \{ -
\frac{1}{\hbar}\int_0^{ \hbar \beta} dt^\prime  \frac{I}{2} \dot 
\varphi^{2} (t^\prime) +
\frac{i \theta }{2 \pi}\int_0^{\hbar \beta} dt^\prime \dot\varphi(t^\prime) -
\frac{1}{\hbar}T[\phi, \varphi] \} \; ,
\end{equation}
where the transformation term,
\begin{equation}
\label{GausTrans}
\exp \{ - \frac{1}{\hbar}T[\phi, \varphi] \} = \prod_t\;\sum_{m_t\in \Z }
\exp \{ - \frac{\hbar a}{I} \frac{\alpha}{2}\; m_t m_t + i m_t [\phi_t
-  \int dt^\prime \Pi(t^\prime - t)  \varphi (t^\prime) ] \; \} \; ,
\end{equation}
is given by an integer Gaussian transform.  The integer sums over
$m_t$ guarantee periodicity for the lattice fields, while the hard
constraint (\ref{RGTcon}) is re-instated mod $2 \pi$ in the limit
$\alpha \rightarrow 0$, as a special case.

By a straightforward series of steps, the exact integration for this
RGT from the continuum to the lattice is performed. The formal steps
are similar to those used for the transfer matrix. To perform the
integral (\ref{eq:FixPoint}), we first expand in Fourier modes,
\begin{equation}
\varphi(t) = \frac{2\pi Q}{\hbar\beta}t + \sum_k \exp\{\frac{i}{\hbar} E_k t\}
\varphi_k \; \ ,
\end{equation}
where $E_k = 2 \pi \hbar k/( a N)$ and $k\in \Z$. The
linear term is present due to non-trivial topological sectors.
Then using the Poisson re-summation formula
to do the sum over $m_t$, we are lead to a closed expression for the perfect 
quantum partition function,
\begin{equation} 
Z(\theta) = 
\int {\cal D}\phi {\cal D}n \ \exp \{ - \frac{1}{\hbar} S[\phi,n]
+ i \theta Q[\phi,n] \} \; .
\end{equation}
The quantum perfect topological charge density is {\em exactly} the same as 
that derived from  the transfer matrix, and is given in 
Eq.(\ref{PerfTopch}). However, the quantum perfect action 
\begin{equation}
\label{nlperfect}
S[\phi,n] = - \frac{I}{ 2 a} \sum_{t=0}^{(N-1)a}
\sum_{j=1}^{\infty} \Delta^{-1}(ja) (\phi_{t+ja} - \phi_t + 2
\pi ( n_{t+a/2 +(j-1)a} + \cdots + n_{t + a/2} )\;)^2 \; ,
\end{equation}
generalizes Eq.(\ref{Villpart3}) to allow for non-local couplings through
$\Delta^{-1}(ja)$, which fall off exponentially in $|ja|$. Further, 
$\phi_t$ and 
$n_{t+a/2}$ are periodic in $t$ by definition. Thus for example
$n_{t+a/2 +(j-1)a} + \cdots + n_{t + a/2}$ represents a ``gauge'' string 
originating at $t+a/2$ and ending at $t+a/2 +(j-1)a$, wrapped around the 
lattice many times as $j$ becomes large. The infinite
lattice propagator $\Delta(t)$ in $E$ space,
\begin{equation}
\label{eq:freeProp}
\tilde \Delta (E) = \sum_{l \in \Z} \frac{\Pi (E +2\pi l\hbar/a)^{2}} {(a
E/\hbar+2\pi l)^2} + \alpha = \frac{1}{\hat E^{2}} - \frac{1}{6}
+ \alpha,
\end{equation}
is given by the standard expression for a perfect lattice scalar
field (see Ref.~\cite{BeWi}) where $\Pi (E) = \hat E/E$, with $\hat E
= (2 \hbar/a) \sin (aE/2 \hbar)$ being the $E$ space Haar function.  

By inspection we see that for $\alpha = 1/6$, $\Delta(t)$ is the standard 
nearest neighbor lattice propagator and thus the action defined in 
Eq.(\ref{nlperfect}) becomes ultra-local. Further, this
is exactly the quantum perfect action, given in Eq.(\ref{Villpart3}), 
derived earlier by the transfer matrix method. Consequently, the notion
of a quantum perfect charge is unchanged in the standard RGT approach,
even with an infinite number of couplings in the perfect action.  


\noindent{\bf Conclusions}

We have presented the concept of a quantum perfect topological charge on
the lattice, and we have worked it out explicitly for the simple case
of a quantum rotor.  We found the perfect action to be of the Villain
type.  This formulation is valid and exact for arbitrarily rough
configurations.

Actions of the Villain type occur generally if one maps non-compact
continuum fields onto compact lattice variables.  They are applicable for
the projection of non-compact continuum gauge fields to compact
lattice link variables \cite{prep}. We also discussed the artifacts in
the classically perfect formulation, where the charge coincides with
the geometric charge, while the classically perfect action turns out
to be the Manton action. These identifications are specific to the
simple, exactly solvable model considered here. The artifacts in the
classically perfect formulation are exponentially suppressed. They are
much smaller than the artifacts in the standard formulation, which
disappear only power-like close to the continuum limit.

\end{document}